\documentclass{article}
\usepackage[dvipdfmx]{graphicx}
\usepackage{amsmath}
\usepackage{ascmac}
\usepackage{fancybox}
\usepackage{amssymb}
\usepackage{amsthm}
\usepackage[cmintegrals]{newtxmath}
\usepackage{algorithmic,algorithm}
\usepackage[top=30truemm,bottom=30truemm,left=25truemm,right=25truemm]{geometry}

\newcommand{\eqdef}{\ensuremath{\stackrel{\mathrm{def}}{=}}}

\rm

\title{Upper Bound on Normalized Maximum Likelihood Codes for Gaussian Mixture Models}

\author{So Hirai and Kenji Yamanishi
        \thanks{
        S. Hirai and K. Yamanishi are with
        Graduate School of Information Science and Technology, The University of Tokyo,
        7-3-1 Hongo, Bunkyo-ku, Tokyo, JAPAN \protect \\
        E-mail: so\_hirai@mist.i.u-tokyo.ac.jp \protect \\
        E-mail: yamanishi@mist.i.u-tokyo.ac.jp
        }}
\date{}

\begin{document}

  \maketitle

  \begin{abstract}
    This paper shows that the normalized maximum likelihood~(NML) code-length calculated in \cite{HiraiNMLjournal} is
    an upper bound on the NML code-length strictly calculated for the Gaussian Mixture Model.
    When we use this upper bound on the NML code-length, we must change the scale of the data sequence to satisfy the restricted domain.
    However,
    we also show that the algorithm for model selection is essentially universal, regardless of the scale conversion of the data in Gaussian Mixture Models,
    and that, consequently, the experimental results in \cite{HiraiNMLjournal} can be used as they are.
    In addition to this, we correct the NML code-length in \cite{HiraiNMLjournal} for generalized logistic distributions.
  \end{abstract}

  \section{Problem Setting}
    In this paper, we consider the problem of model selection in which we aim to calculate the number of clusters for a Gaussian Mixture Model (GMM).
    Let us use the given sequence ${\bf x}^n=({\bf x}_1,\cdots ,{\bf x}_n),\ {\bf x}_i=(x_{i1},\cdots ,x_{im})^{\top} \ (i=1,\cdots ,n)$.
    Here, we use the Gaussian Model Class ${\cal N}(\mu, \Sigma),\ {\bf \mu}\in {\mathbb R}^m,\Sigma \in {\mathbb R}^{m\times m}$,
    and calculate the normalized maximum likelihood (NML) code-length for the Gaussian Model.
    The Gaussian distribution for data sequence $\mathbf{x}^{n}$ is defined as follows:
    \begin{eqnarray*}
      f({\bf x}^n; \mu , \Sigma )
      = \frac{1}{(2\pi )^{\frac{mn}{2}}\cdot |\Sigma |^{\frac{n}{2}}}\exp \Big
      \{ -\frac{1}{2} \sum _{i=1}^{n} ({\bf x}_i-{\bf \mu})^{\top} \Sigma^{-1} ({\bf x}_i-{\bf \mu}) \Big \} . \label{eqn:prob_gaussian}
    \end{eqnarray*}
    We define the NML distribution $f_{\mathrm{NML}}$ relative to a model class $\mathcal{M}=\{ f(X^n;\theta):\theta \in \Theta \}$ by
    \begin{equation}
      f_{\mathrm{NML}}({\bf x}^n;\mathcal{M})\buildrel \rm def \over =\frac{f({\bf x}^n ; \hat{\mu}({\bf x}^n),\hat{\Sigma }({\bf x}^n))}
      {\int_{Y} f({\bf y}^n ; \hat{\mu}({\bf y}^n),\hat{\Sigma }({\bf y}^n)){\rm d}{\bf y}^n }.  \label{eqn:NMLdistributionGau}
    \end{equation}
    Here, $Y$ is the restricted domain for $\mathbf{x}^n$.
    By using this restriction, we can calculate the NML code-length without divergence.

    The NML code-length for the GMM is defined as follows with the latent variable $z^n$:
    \begin{eqnarray}
      L_{\mathrm{NML}}({\bf x}^n,z^n;Y,\mathcal{M}(K))
      &\eqdef& -\log f_{\mathrm{NML}}({\bf x}^n,z^n) \notag \\
      &=& -\log f({\bf x}^n,z^n ; {\cal M}(K), \hat{\theta}({\bf x}^n,z^n))
       + \log \mathcal{C}(\mathcal{M}(K),n), \label{eqn:NMLcodeGauMix} \\
      \mathcal{C}(\mathcal{M}(K),n)
      &=& \sum_{w^n} \int_{Y} f({\bf y}^n,w^n ; {\cal M}(K), \hat{\theta}({\bf y}^n,w^n)){\rm d}{\bf y}^n .
    \end{eqnarray}
    Here, $\theta=(\pi,\mu,\Sigma)$ is the set of parameters.
    We consider the problem of model selection for the GMM using (\ref{eqn:NMLcodeGauMix}) as a criterion.

  \section{Influence of scale conversion of data on model selection}
  \label{sec:scale_change}
    When we use the NML code-length defined by (\ref{eqn:NMLcodeGauMix}),
    we must change the scale of the data sequence to satisfy the restricted domain $Y$ (e.g., to multiply $1/\alpha$, etc.).
    In this section, we show that the model selection algorithm is essentially universal, regardless of the scale conversion of the data.

    Let us consider the NML code-length for a GMM as $L_{\mathrm{NML}}(\mathbf{x}^n,z^n;Y,\mathcal{M}(K))$.
    We can derive the definition of the code-length from (\ref{eqn:NMLcodeGauMix}).
    The term influenced by the scale of the data is the first term in Eq. (\ref{eqn:NMLcodeGauMix}).
    Here, in order to evaluate the influence of the first term on the scale conversion of the data,
    we used the dataset $\mathbf{x}_{\alpha}^n \ (=\mathbf{x}^n/\alpha)$, which we calculated by multiplying $\mathbf{x}^n$ by $1/\alpha$.
    We considered model selection when either $\mathbf{x}^n$ or $\mathbf{x}_{\alpha}^n$ was used, and evaluated the difference between them.
    Because it is important for model selection to evaluate the difference between $\mathcal{M}(K_1)$ and $\mathcal{M}(K_2)$,
    we focused on the difference in the first term of Eq. (\ref{eqn:NMLcodeGauMix}):
    \begin{eqnarray}
      && -\log f(\mathbf{x}_{\alpha}^n,z^n ; {\cal M}(K_1), \hat{\theta}(\mathbf{x}_{\alpha}^n,z^n))
      -( -\log f(\mathbf{x}_{\alpha}^n,z^n ; {\cal M}(K_2), \hat{\theta}(\mathbf{x}_{\alpha}^n,z^n)) \notag \\
      &=& C + \sum_{k=1}^{K_1} \sum_{j=1}^m \frac{h_k}{2}\log \hat{\lambda}_j (\mathbf{x}_{\alpha}^n)
      - \sum_{k=1}^{K_2} \sum_{j=1}^m \frac{h'_k}{2}\log \hat{\lambda}'_j (\mathbf{x}_{\alpha}^n) \notag \\
      &=& C + \sum_{k=1}^{K_1} \sum_{j=1}^m \frac{h_k}{2} \left\{ \log \hat{\lambda}_j (\mathbf{x}^n) - 2\log \alpha \right\}
       -\sum_{k=1}^{K_2} \sum_{j=1}^m \frac{h'_k}{2} \left\{ \log \hat{\lambda}'_j (\mathbf{x}^n) - 2\log \alpha \right \} \\
      &=& C + \sum_{k=1}^{K_1} \sum_{j=1}^m \frac{h_k}{2}\log \hat{\lambda}_j (\mathbf{x}^n)
      - \sum_{k=1}^{K_2} \sum_{j=1}^m \frac{h'_k}{2}\log \hat{\lambda}'_j (\mathbf{x}^n) \notag \\
      &=& -\log f(\mathbf{x}^n,z^n ; {\cal M}(K_1), \hat{\theta}(\mathbf{x}^n,z^n))
      -( -\log f(\mathbf{x}^n,z^n ; {\cal M}(K_2), \hat{\theta}(\mathbf{x}^n,z^n))
    \end{eqnarray}
    where we define
    \begin{eqnarray*}
      C \eqdef -\sum_{k=1}^{K_1} h_k\log \hat{\pi}_k + \sum_{k=1}^{K_2} h'_k\log \hat{\pi}'_k
      + \sum_{k=1}^{K_1}\frac{m{h_k}}{2}\log 2\pi \mathrm{e} - \sum_{k=1}^{K_2}\frac{m{h'_k}}{2}\log 2\pi' \mathrm{e}
    \end{eqnarray*}
    and each $h_k,h'_k$ represents the number of data that belong to $k$ under model class $\mathcal{M}(K_1),\mathcal{M}(K_2)$.
    This shows us that the difference in code-length is unaffected by the scale conversion of the data.
    Consequently, the data can be processed such that it satisfies the restriction and can be used for model selection without changing the result.

    In what follows, we define the restricted domain with the maximum likelihood estimator~(MLE) for parameter $\hat{\mu},\hat{\lambda}=(\hat{\lambda}_1,\cdots,\hat{\lambda}_m)$,
    where each $\hat{\lambda}_j$ is a $j$-th eigenvalue of $\hat{\Sigma}$.
    These MLEs are changed by scale conversion as follows:
    \begin{eqnarray}
      \hat{\mu}_{\alpha} &=& \frac{1}{n} \sum_{i=1}^n x_{i\alpha}
      = \frac{1}{n\alpha} \sum_{i=1}^n x_{i} = \frac{1}{\alpha} \hat{\mu}, \label{eqn:scale_mu} \\
      \hat{\Sigma}_{\alpha} &=& \frac{1}{n} \sum_{i=1}^n (x_{i\alpha}-\hat{\mu}_{\alpha}) (x_{i\alpha}-\hat{\mu}_{\alpha})^{\top}
      = \frac{1}{n\alpha^2} \sum_{i=1}^n (x_{i}-\hat{\mu}) (x_{i}-\hat{\mu})^{\top} = \frac{1}{\alpha^2}\hat{\Sigma}
      = \frac{1}{\alpha^2} U\Lambda U^{\top}\notag \\
      \Leftrightarrow \hat{\lambda}_{j\alpha} &=& \frac{1}{\alpha^2} \hat{\lambda}_j. \label{eqn:scale_Sigma}
    \end{eqnarray}
    Here, the data sequence $\mathbf{x}^n$ denotes data assigned to a cluster.
    This shows that the MLEs of the parameters can be converted to an arbitrary size by scale conversion.

  \section{Upper bound on the NML code-length}
    As explained in Section \ref{sec:scale_change}, the model selection algorithm is essentially universal, regardless of the scale conversion of the data.
    Thus, we can restrict the domain to calculate the NML code-length.
    We here demonstrate that the code-length calculated by \cite{HiraiNMLjournal} is an upper bound on the NML code-length strictly calculated for the GMM.
    Consequently, this upper bound to the NML code-length (hereafter referred to as the uNML code-length) can be used for model selection.

    \subsection{1st modification to \cite{HiraiNMLjournal}}
      We use Shtarkov's minmax regret with a restricted domain as follows:
      \begin{eqnarray}
        && \min _Q \max _{{\bf x}^n \in Y(R, \epsilon_1,\epsilon_2)} \Big \{ -\log Q({\bf x}^n) -\min _{\theta}(-\log P({\bf x}^n| {\theta})) \Big \} \label{eqn:agenda_minmaxY} \\
        Y(R, \epsilon_1,\epsilon_2) &\eqdef& \{ {\bf y}^n| \ ||\hat{\mu}(\mathbf{y}^n)||^2 \leq R,
        \epsilon_{1j} \leq \hat{\lambda}_j({\bf y}^n) \leq \epsilon_{2j} \leq \epsilon_{2} < 1 \ (j=1,\cdots ,m) , \notag \\
        && \qquad \qquad \frac{\mathrm{Vol}(\mathcal{O}(m))}{2^m} \cdot \epsilon_{2}^{\frac{m(m-1)}{2}} \leq 1,\ {\bf y}^n \in {\cal X}^n  \}. \label{eqn:agenda_rangeY}
      \end{eqnarray}

    \subsection{2nd modification to \cite{HiraiNMLjournal}}
      Using the restricted domain (\ref{eqn:agenda_rangeY}), we can calculate an upper bound on NML code-length as follows:
      \begin{eqnarray}
        {\cal C}(\mathcal{M},n)
        &=& \int_{Y(R, \epsilon_1,\epsilon_2)} f({\bf y}^n ; \hat{\mu}({\bf y}^n),\hat{\Sigma }({\bf y}^n)) \ \mathrm{d}\mathbf{y}^n \notag \\
        &=& \int \delta (\hat{\mu}(\mathbf{x}^n)=\hat{\mu},\hat{\Sigma}(\mathbf{x}^n)=\hat{\Sigma}) \ \mathrm{d}\mathbf{y}^n \
        \int_{||\hat{\mu}||^2 \leq R} \ \mathrm{d}\mu \ \int g(\hat{\lambda}) \ {\rm d}\hat{\Sigma} \notag \\
        &=& \int_{||\hat{\mu}||^2 \leq R} \ \mathrm{d}\mu \ \int \mathrm{d}U
        \ \int_{\epsilon_{1j} \leq \hat{\lambda}_j \leq \epsilon_{2j}} \
        g(\hat{\lambda}) \prod_{1\leq i < j \leq m} | \hat{\lambda}_i - \hat{\lambda}_j | \ {\rm d}\hat{\lambda} \label{eqn:agenda_orthogonal_integral} \\
        &<& \int \mathrm{d}U \cdot \epsilon_{2}^{\frac{m(m-1)}{2}} \ \int_{||\hat{\mu}||^2 \leq R} \ \mathrm{d}\mu \
        \int_{\epsilon_{1j} \leq \hat{\lambda}_j \leq \epsilon_{2j}} g(\hat{\lambda}) \ {\rm d}\hat{\lambda} \notag \\
        &=& \frac{\mathrm{Vol}(\mathcal{O}(m))}{2^m} \cdot \epsilon_{2}^{\frac{m(m-1)}{2}} \cdot
        \frac{2^{m+1}R^{\frac{m}{2}} (\prod_{j=1}^m {\epsilon_{1j}}^{-\frac{m}{2}} - \prod_{j=1}^m {\epsilon_{2j}}^{-\frac{m}{2}}) } {m^{m+1}\cdot \Gamma \left( \frac{m}{2} \right)}
        \cdot \left( \frac{n}{2{\rm e}} \right)^{\frac{mn}{2}} \frac{1}{\Gamma_ m(\frac{n-1}{2})}  \notag \\
        &<& \frac{2^{m+1}R^{\frac{m}{2}} \prod_{j=1}^m {\epsilon_{1j}}^{-\frac{m}{2}} } {m^{m+1}\cdot \Gamma \left( \frac{m}{2} \right)}
        \times \left( \frac{n}{2{\rm e}} \right)^{\frac{mn}{2}} \frac{1}{\Gamma_ m(\frac{n-1}{2})}  \notag \\
        &=& B(m,R,\epsilon) \times \left( \frac{n}{2{\rm e}} \right)^{\frac{mn}{2}} \frac{1}{\Gamma_ m(\frac{n-1}{2})}.\label{eqn:agenda_paraComp}
      \end{eqnarray}
      Here, Eq. (\ref{eqn:agenda_paraComp}) is identical to the code-length calculated in \cite{HiraiNMLjournal}
      and represents an upper bound on the NML code-length.
      Given this conclusion, we can use the uNML (\ref{eqn:agenda_paraComp}) for model selection with data that satisfies the restricted domain (\ref{eqn:agenda_rangeY}).

      Here, we define a uNML code-length as follows:
      \begin{eqnarray*}
        L_{\mathrm{uNML}}({\bf x}^n;Y,\mathcal{M})
        &\eqdef& -\log f_{\mathrm{uNML}}({\bf x}^n;\mathcal{M}) \notag \\
        &=& -\log f({\bf x}^n; \mathcal{M}, \hat{\theta}({\bf x}^n))
         + \log \mathcal{C}_{\mathrm{u}}(\mathcal{M},n), \\
        \mathcal{C}_{\mathrm{u}}(\mathcal{M},n)
        &=& B(m,R,\epsilon) \times \left( \frac{n}{2{\rm e}} \right)^{\frac{mn}{2}} \frac{1}{\Gamma_ m(\frac{n-1}{2})} .
      \end{eqnarray*}
      By calculating the uNML code-length for GMMs, we can use this definition of the normalization term in the same manner as \cite{HiraiNMLjournal}.

    \subsection{Handling this code-length}
      Using this uNML code-length, we have to change the scale of the data sequence to satisfy the restricted domain (\ref{eqn:agenda_rangeY}).
      From the discussion in Section \ref{sec:scale_change}, we can use it in computation for model selection without the need for the scale conversion of the data.
      In the experiments described in \cite{HiraiNMLjournal}, the artificial dataset was scaled to satisfy the restricted domain (\ref{eqn:agenda_rangeY}).
      Therefore, the experimental results in \cite{HiraiNMLjournal} can be used as they are.

  \section{Summary of modifications to \cite{HiraiNMLjournal}}
    The following is a modification of \cite{HiraiNMLjournal} that describes calculating an upper bound on the NML code-length.

    Let an observed data sequence be ${\bf x}^n=({\bf x}_1,\cdots ,{\bf x}_n)$ where ${\bf x}_i=(x_{i1},\cdots ,x_{im})^{\top} \ (i=1,\cdots ,n)$.
    We use a class of Gaussian distributions: ${\cal N}(\mu, \Sigma)$, where ${\bf \mu}\in {\mathbb R}^m$ is a mean vector,
    $\Sigma \in {\mathbb R}^{m\times m}$ is a covariance matrix, and $m$ is the dimension of ${\bf x}_i.$
    A probability density function of $\mathbf{x}^{n}$ for the Gaussian distribution is given by
    \begin{eqnarray*}
      f({\bf x}^n; \mu , \Sigma )
      = \frac{1}{(2\pi )^{\frac{mn}{2}}\cdot |\Sigma |^{\frac{n}{2}}}\exp \Big
      \{ -\frac{1}{2} \sum _{i=1}^{n} ({\bf x}_i-{\bf \mu})^{\top} \Sigma^{-1} ({\bf x}_i-{\bf \mu}) \Big \}, \label{eqn:gaussian}
    \end{eqnarray*}
    and the NML distribution based on the Gaussian distribution is defined as follows:
    \begin{equation}
      f_{\mathrm{NML}}({\bf x}^n)\buildrel \rm def \over =\frac{f({\bf x}^n ; \hat{\mu}({\bf x}^n),\hat{\Sigma }({\bf x}^n))}
      {\int_{Y(R, \epsilon_1,\epsilon_2)} f({\bf y}^n ; \hat{\mu}({\bf y}^n),\hat{\Sigma }({\bf y}^n)){\rm d}{\bf y}^n }
    \end{equation}
    where $\hat{\mu}({\bf x}^n)$ and $\hat{\Sigma}({\bf x}^n)$ are the MLEs of $\mu$ and $\Sigma$ respectively:
    \begin{eqnarray*}
      \hat{\mu}({\bf x}^n)&=&\frac{1}{n}\sum^n_{i=1}{\bf x}_i,\\
      \hat{\Sigma}({\bf x}^n)&=&\frac{1}{n}\sum^{n}_{i=1}({\bf x}_i-\hat{\mu}({\bf x}^n))({\bf x}_i-\hat{\mu}({\bf x}^n))^{\top}.
    \end{eqnarray*}

    For given constants $R,\epsilon_1,\epsilon_2$, we set a restricted domain as follows:
    \begin{eqnarray}
      Y(R, \epsilon_1,\epsilon_2) &\eqdef& \{ {\bf y}^n| \ ||\hat{\mu}(\mathbf{y}^n)||^2 \leq R,
      \epsilon_{1j} \leq \hat{\lambda}_j({\bf y}^n) \leq \epsilon_{2j} \leq \epsilon_{2} < 1 \ (j=1,\cdots ,m) , \notag \\
      && \qquad \qquad \frac{\mathrm{Vol}(\mathcal{O}(m))}{2^m} \cdot \epsilon_{2}^{\frac{m(m-1)}{2}} \leq 1,\ {\bf y}^n \in {\cal X}^n  \}, \label{eqn:rangeY}
    \end{eqnarray}
    where $\hat{\lambda}_j({\bf y}^n) \ (j=1,\cdots ,m)$ are eigenvalues of $\hat{\Sigma}({\bf y}^n)$.
    This restriction facilitates the calculation of an upper bound on normalization term ${\cal C}({\cal M},n)$, as shown below.

    First, by substituting MLE $\hat{\mu}({\bf x}^n),\ \hat{\Sigma}({\bf x}^n)$ into Eq. (\ref{eqn:gaussian}), the numerator of Eq. (\ref{eqn:NMLdistributionGau}) can be expressed as follows:
    \begin{eqnarray}
      f({\bf x}^n ; \hat{\mu}({\bf x}^n),\hat{\Sigma }({\bf x}^n))
      &=& \prod _{i=1}^n \frac{1}{(2\pi )^{\frac{m}{2}} |\hat{\Sigma}({\bf x}^n) |^{\frac{1}{2}}}
      \times \exp \Big \{ -\frac{1}{2} ({\bf x}_i-\hat{\mu}({\bf x}^n))^{\top} \hat{\Sigma}({\bf x}^n) ^{-1} ({\bf x}_i-\hat{\mu}({\bf x}^n))  \Big \} \\
      &=& (2\pi \mathrm{e})^{-\frac{mn}{2}} \prod_{j=1}^m \hat{\lambda}_j(\mathbf{x}^n)^{-\frac{n}{2}}.  \label{eqn:NMLNumerator}
    \end{eqnarray}

    Next, we calculate the denominator in Eq. (\ref{eqn:NMLdistributionGau}).
    Using the fact that $\hat{\mu}({\bf x}^n)$ and $\hat{\Sigma}({\bf x}^n)$ are sufficient statistics,
    we can calculate the normalization term as an integral with respect to $\hat{\mu},\hat{\Sigma}$.
    Because MLEs are sufficient statistics, $f({\bf x}^n;\mu ,\Sigma )$ is decomposed as follows:
    \begin{equation*}
      f({\bf x}^n;\mu ,\Sigma )=f({\bf x}^n | \hat{\mu }({\bf x}^n),\hat{\Sigma}({\bf x}^n))\cdot
      g_1(\hat{\mu }({\bf x}^n);\mu ,\Sigma )\cdot g_2(\hat{\Sigma}({\bf x}^n);\Sigma ).
    \end{equation*}
    where
    \begin{eqnarray*}
      g_1(\hat{\mu }({\bf x}^n);\mu ,\Sigma )
      &\eqdef& \frac{1}{(2\pi / n)^{\frac{m}{2}}|\Sigma |^{\frac{1}{2}}}
      \exp \Big \{ -\frac{1}{2/n} (\hat{\mu}({\bf x}^n)-{\bf \mu})^{\top} \Sigma^{-1} (\hat{\mu}({\bf x}^n)-{\bf \mu}) \Big \}, \notag \\
      g_2(\hat{\Sigma}({\bf x}^n);\Sigma )
      &\eqdef& \frac{|\hat{\Sigma }({\bf x}^n)|^{\frac{n-m-2}{2}} }
      { 2^{\frac{m(n-1)}{2}} |\frac{1}{n} \Sigma |^{\frac{n-1}{2}} \Gamma _m(\frac{n-1}{2})}
       \times \exp \Big \{ -\frac{1}{2} {\rm Tr} (n\Sigma^{-1} \hat{\Sigma}({\bf x}^n))\Big \} .
    \end{eqnarray*}
    Here, we define the function $f({\bf x}^n | \hat{\mu }({\bf x}^n),\hat{\Sigma}({\bf x}^n))=
    \delta (\hat{\mu}(\mathbf{x}^n)=\hat{\mu},\hat{\Sigma}(\mathbf{x}^n)=\hat{\Sigma})$.
    We fix values $\hat{\mu}(\mathbf{x}^n)=\hat{\mu},\hat{\Sigma}(\mathbf{x}^n)=\hat{\Sigma}$, and let
    \begin{eqnarray}
      g(\hat{\lambda}) &{\buildrel \rm def \over {=}}& g_1(\hat{\mu };\hat{\mu },\hat{\Sigma })\cdot g_2(\hat{\Sigma};\hat{\Sigma }) \\
      &=& \frac{n^{\frac{mn}{2}}} {2^{\frac{mn}{2}}\pi^{\frac{m}{2}}{\rm e}^{\frac{mn}{2}} \Gamma_m(\frac{n-1}{2})}
      \cdot \prod_{j=1}^m {\hat{\lambda}_j}^{-\frac{m}{2}-1}.
    \end{eqnarray}
    By letting the normalization term in Eq. (\ref{eqn:NMLdistributionGau}) be ${\cal C}({\cal M},n)$,
    we can calculate an upper bound on it by integrating $g(\hat{\lambda})$ with respect to $\hat{\mu},\hat{\Sigma}$ over the restricted domain as follows:
    \begin{eqnarray}
      {\cal C}(\mathcal{M},n)
      &=& \int_{Y(R, \epsilon_1,\epsilon_2)} f({\bf y}^n ; \hat{\mu}({\bf y}^n),\hat{\Sigma }({\bf y}^n)) \ \mathrm{d}\mathbf{y}^n \notag \\
      &=& \int \delta (\hat{\mu}(\mathbf{x}^n)=\hat{\mu},\hat{\Sigma}(\mathbf{x}^n)=\hat{\Sigma}) \ \mathrm{d}\mathbf{y}^n \
      \int_{||\hat{\mu}||^2 \leq R} \ \mathrm{d}\mu \ \int g(\hat{\lambda}) \ {\rm d}\hat{\Sigma} \notag \\
      &=& \int_{||\hat{\mu}||^2 \leq R} \ \mathrm{d}\mu \ \int \mathrm{d}U
      \ \int_{\epsilon_{1j} \leq \hat{\lambda}_j \leq \epsilon_{2j}} \
      g(\hat{\lambda}) \prod_{1\leq i < j \leq m} | \hat{\lambda}_i - \hat{\lambda}_j | \ {\rm d}\hat{\lambda} \label{eqn:orthogonal_integral} \\
      &<& \int \mathrm{d}U \cdot \epsilon_{2}^{\frac{m(m-1)}{2}} \ \int_{||\hat{\mu}||^2 \leq R} \ \mathrm{d}\mu \
      \int_{\epsilon_{1j} \leq \hat{\lambda}_j \leq \epsilon_{2j}} g(\hat{\lambda}) \ {\rm d}\hat{\lambda} \notag \\
      &=& \frac{\mathrm{Vol}(\mathcal{O}(m))}{2^m} \cdot \epsilon_{2}^{\frac{m(m-1)}{2}} \cdot
      \frac{2^{m+1}R^{\frac{m}{2}} (\prod_{j=1}^m {\epsilon_{1j}}^{-\frac{m}{2}} - \prod_{j=1}^m {\epsilon_{2j}}^{-\frac{m}{2}}) } {m^{m+1}\cdot \Gamma \left( \frac{m}{2} \right)}
      \cdot \left( \frac{n}{2{\rm e}} \right)^{\frac{mn}{2}} \frac{1}{\Gamma_ m(\frac{n-1}{2})}  \notag \\
      &<& \frac{2^{m+1}R^{\frac{m}{2}} \prod_{j=1}^m {\epsilon_{1j}}^{-\frac{m}{2}} } {m^{m+1}\cdot \Gamma \left( \frac{m}{2} \right)}
      \times \left( \frac{n}{2{\rm e}} \right)^{\frac{mn}{2}} \frac{1}{\Gamma_ m(\frac{n-1}{2})}  \notag \\
      &=& B(m,R,\epsilon) \times \left( \frac{n}{2{\rm e}} \right)^{\frac{mn}{2}} \frac{1}{\Gamma_ m(\frac{n-1}{2})}.\label{eqn:paraComp}
    \end{eqnarray}
    where Eq. (\ref{eqn:orthogonal_integral}) is given in \cite{mathai1997jacobians}, and we define $B(m,R,\epsilon)$ by
    \begin{equation*}
      B(m,R,\epsilon)\ {\buildrel \rm def \over {=}}\ \frac{2^{m+1}R^{\frac{m}{2}} \prod_{j=1}^m {\epsilon_{1j}}^{-\frac{m}{2}}} {m^{m+1}\cdot \Gamma \left( \frac{m}{2} \right)}.
    \end{equation*}
    $B(m,R,\epsilon)$ does not depend on a number of data $n$.
    Because (\ref{eqn:paraComp}) is finite, an upper bound on the normalization term ${\cal C}({\cal M},n)$ does not diverge.

    Here, we can define an upper bound on NML (uNML) code-length as follows:
    \begin{eqnarray*}
      L_{\mathrm{uNML}}({\bf x}^n;Y,\mathcal{M})
      &\eqdef& -\log f_{\mathrm{uNML}}({\bf x}^n;\mathcal{M}) \notag \\
      &=& -\log f({\bf x}^n; \mathcal{M}, \hat{\theta}({\bf x}^n))
       + \log \mathcal{C}_{\mathrm{u}}(\mathcal{M},n), \label{eqn:uNMLcodeGauMix} \\
      \mathcal{C}_{\mathrm{u}}(\mathcal{M},n)
      &=& B(m,R,\epsilon) \times \left( \frac{n}{2{\rm e}} \right)^{\frac{mn}{2}} \frac{1}{\Gamma_ m(\frac{n-1}{2})} .
    \end{eqnarray*}
    In calculating the uNML code-length for GMMs, we can use this definition of the normalization term in the same manner as \cite{HiraiNMLjournal}.

  \section{Correcting the NML for generalized logistic distributions}
    The following is a modification to \cite{HiraiNMLjournal} that describes the correction to the NML for generalized logistic distributions.

    In \cite{HiraiNMLjournal}, the generalized logistic distribution is used as an example of the exponential family.
    The density function of $x^n$ for a generalized logistic distribution with parameter $\theta $ is defined as
    \begin{equation*}
    f(x^n;\theta) = \prod_{i=1}^n \frac{\theta {\rm e}^{-x_i} }{(1+{\rm e}^{-x_i})^{\theta+1}}.
    \end{equation*}
    The MLE of $\theta$ is analytically obtained as
    $\hat{\theta}(x^n) = n/(\sum_{i=1}^n \log (1+{\rm e}^{-x_i}))$.
    Thus, the joint density of $x^n$ is written as
    \begin{eqnarray*}
    f(x^n;\theta) &=& \theta ^n \cdot \exp \left\{ -\sum_{i=1}^n x_i - \frac{n(\theta+1)}{\hat{\theta}(x^n)} \right\} \\
    &=& H ( x^n | \hat{\theta}(x^n) ) \cdot g(\hat{\theta}(x^n); \theta ),
    \end{eqnarray*}
    where $n/\hat{\theta}(x^n)$ is distributed according to the Gamma distribution with a shape parameter $n$ and a scale parameter $1/\theta$.

    Here, we correct the function $g(\hat{\theta}(x^n); \theta )$, changing the result of the NML code-length.
    First, the function $g(\hat{\theta}(x^n); \theta )$ is written as
    \begin{eqnarray*}
      g(\hat{\theta}(x^n);\theta)
      = \frac{\theta^n}{(n-1)!}\cdot \frac{n^n}{{\hat{\theta}(x^n)}^{n+1}}
    \cdot \exp \left\{ -\theta \cdot \frac{n}{\hat{\theta}(x^n)} \right\}.
    \end{eqnarray*}
    By fixing $\hat{\theta}(x^n)=\hat{\theta}$, we have
    \begin{equation*}
    g(\hat{\theta };\hat{\theta}) = \frac{n^n}{{\rm e}^n (n-1)!} \cdot \frac{1}{\hat{\theta}}.
    \end{equation*}
    Then, the normalization term ${\cal C}({\cal M})$
    is calculated by taking an integral of $g(\hat{\theta};\hat{\theta})$
    with respect to $\hat{\theta}$. Here, we use hyper-parameters $\theta_{\min},\theta_{\max}$ to restrict
    the domain for the integral to be taken as follows:
    \begin{equation*}
    Y(\theta_{\min},\theta_{\max}) = \left\{ y^n | \theta_{\min} \leq \hat{\theta}(y^n) \leq \theta_{\max} \right\}.
    \end{equation*}
    Then, we have
    \begin{eqnarray*}
    {\cal C}({\cal M}) &=& \int_{Y(\theta_{\min},\theta_{\max})} g(\hat{\theta };\hat{\theta}) {\rm d}\hat{\theta} \\
    &=& \frac{n^n}{{\rm e}^n (n-1)!} \int_{\theta_{\min}}^{\theta_{\max}} \frac{1}{\hat{\theta}}\ {\rm d}\hat{\theta} \\
    &=& \frac{n^n}{{\rm e}^n (n-1)!} \cdot \log \frac{\theta_{\max}}{\theta_{\min}}.
    \end{eqnarray*}
    Hence, we obtain an approximation of the normalization term ${\cal C}({\cal M})$ for generalized logistic distributions in an analytical manner.

  \section{Acknowledgements}
    This work was supported by JST CREST No. JP- MJCR1304.
    We thank Professor Jun'ichi Takeuchi at Kyushu University and Mr. Kohei Miyaguchi at The University of Tokyo for helpful discussion.

  \rm

\end{document}